\documentclass[preprint,11pt,times]{elsarticle}

\usepackage{ifpdf}
\ifpdf
  \usepackage[1.7]{bxpdfver}
\fi
\usepackage{hyperref}
\usepackage{amssymb}
\usepackage{amsmath}
\usepackage{amsfonts}
\usepackage{multirow}

\journal{arXiv}
\begin{document}

\begin{frontmatter}

\title{Fast Magnetic Resonance Simulation Using Combined Update with Grouped Isochromats}
\author[First]{Hidenori Takeshima}
%\affiliation[First]{organization={Advanced Technology Research Department, Research and Development Center,\\
%            Canon Medical Systems Corporation},
%            %addressline={},
%            city={Tokyo},
%            %postcode={},
%            %state={},
%            country={Japan}}

\address[First]{Advanced Technology Research Department, Research and Development Center,\\
                Canon Medical Systems Corporation, Tokyo, Japan}

\begin{abstract}

This work aims to overcome an assumption of conventional MR simulators:
Individual isochromats should be simulated individually.
To reduce the computational times of MR simulation,
a new simulation method using grouped isochromats is proposed.
When multiple isochromats are grouped before simulations,
some parts of the simulation can be shared in each group.
For a certain gradient type, the isochromats in the group can be easily chosen
for ensuring that they behave the same.
For example, the group can be defined as the isochromats whose locations along x-axis,
T1, T2 and magnetic field inhomogeneity values are the same values.
In such groups, simulations can be combined when a pulse sequence with
the magnetic field gradient along x-axis only are processed.
The processing times of the conventional and proposed methods were evaluated with
several sequences including fast spin echo (FSE) and echo-planar imaging (EPI) sequences.
The simulation times of the proposed method were 3 to 72 times faster than those of the conventional methods.
In the cases of 27.5 million isochromats using single instruction multiple data (SIMD) instructions and multi-threading,
the conventional method simulated FSE and EPI sequences in 208.4 and 66.4 seconds, respectively.
In the same cases, the proposed method simulated these sequences in 38.1 and 7.1 seconds, respectively.

\end{abstract}


\end{frontmatter}

\section{Introduction}

MR simulation is an important part of the development, prototyping, optimization,
and evaluation of various aspects of MR. There are many implementations
\cite{Taniguchi1,Taniguchi2,SIMRI,Jochimsen,Latta,JEMRIS,MRISIMUL,MRiLab,Kose,coreMRI,Scholand,KomaMRI,Takeshima}
for simulating the Bloch equations and their extensions
\cite{Bloch,Torrey,McConnell}.
In MR simulators, the magnetizations of isochromats are updated using small step sizes
in the order of microseconds. Realistic simulations require millions of isochromats.
Even when the spatial resolution of the phantom is not high,
it is known that the number of subvoxels should be increased for suppressing artifacts \cite{Kose}.

A widespread problem of the simulations was to reduce the computational times.
Existing work used algorithm-based and hardware-based methods for reducing the computational times.
The algorithm-based methods reduced the amount of computation.
These methods included utilization of combined transitions \cite{Taniguchi1,Taniguchi2,Scholand,Takeshima},
and no temporal changes \cite{MRISIMUL}. The hardware-based methods used parallel computing.
These methods included utilization of multi-threading \cite{JEMRIS,Kose,Takeshima},
single instruction multiple data (SIMD) \cite{Kose,Takeshima},
general-purpose graphics processing unit (GPGPU) \cite{MRISIMUL,MRiLab,Kose,KomaMRI},
computer clusters \cite{SIMRI,JEMRIS},
and cloud computing \cite{coreMRI}.
These methods could reduce the computational times. However, the computational costs of the MR simulation
were still high when millions of isochromats were used.

The aim of this paper is to accelerate the simulations by overcoming a common assumption of
conventional MR simulators: Individual isochromats should be simulated individually.
Under this assumption, isochromats could be updated simultaneously only if they were computed
in parallel using the hardware-based methods.

To further accelerate the simulations, a new computation method using \linebreak%%%
``grouped isochromats'' is proposed.
When multiple isochromats are grouped before a simulation,
some parts of the simulation can be shared in each group.
For a certain gradient type, the isochromats in the group can be easily chosen for
ensuring that they behave the same. In such gradient types, the multiple isochromats
were combined and updated simultaneously without relying on the hardware-based methods.
The preliminary version of this work was presented as an abstract \cite{Takeshima2}.

\section{Theory}

The theory section consists of four parts: Basics of MR simulation, computational complexities of
conventional methods, efficient simulations of with-RF and with-ADC subsequences, and grouping criteria for the efficient simulations.
The first and second parts explain the computational complexities of conventional methods.
The remaining parts give the new idea and its computational complexities of the proposed method.

\subsection{Basics of MR simulation}%
\begin{figure}[t]%
\centering%
\includegraphics{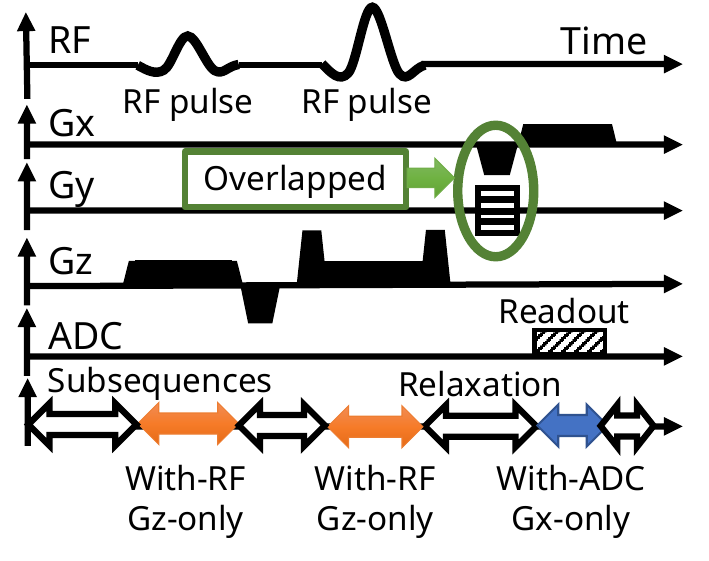}
\caption{Types of subsequences and their groups.
The durations where RF pulses are active are classified as with-RF subsequences.
The duration where analog-to-digital converters are active is classified as a with-ADC subsequence.
Remaining durations are classified as relaxation subsequences.
In this example, only $G_z$ is active in with-RF subsequences,
and only $G_x$ is active in a with-ADC subsequence.
The former and latter subsequences can be considered as Gz-only and Gx-only groups, respectively.
Abbreviations: RF, radiofrequency, ADC, analog-to-digital converter.}\label{fig1}
\end{figure}

Let the direction of the static magnetic field be the z axis.
Let the remaining axes representing the rotating frame around the z axis be the x and y axes.
Let the time be t. Let the constant $ \gamma $  be the gyromagnetic ratio.
In the cases of the Bloch equations \cite{Bloch}, the magnetization vector
$ M(k,t)=(M_x (k,t),M_y (k,t),M_z (k,t))^T $ of the $k$-th isochromat ($ k=1,\ldots,K $) satisfies
\begin{equation}\label{eq1}
\frac{dM(k,t)}{dt}=\gamma (M(k,t) \times B(k,t))+C(k)(M_{static}(k) - M(k,t))
\end{equation}
where the static magnetization vector $M_{static} (k)=(0,0,M_0 (k))^T$ and
the matrix $C(k)=diag(-1/T_2 (k),-1/T_2 (k),-1/T_1 (k))$ represent the constants of the $k$-th isochromat,
and the magnetic field vector $B(k,t)=(B_x (k,t),B_y (k,t),B_z (k,t))^T$ represent a pulse sequence
to be simulated at the $k$-th isochromat.
In the pulse sequence, $B_1 (k,t)=B_x (k,t)+iB_y (k,t)$ represents the radiofrequency (RF) pulses,
$B_z (k,t)=G(t)r(k,t)+\Delta B_0 (k)$ represents the magnetic fields from the gradient fields
$G(t)=(G_x (t),G_y (t),G_z (t))^T$ at the $k$-th location
$r(k,t)=(r_x (k,t),r_y (k,t),r_z (k,t))^T$ and inhomogeneity $\Delta B_0 (k)$.
In the cases of static isochromats, $r(k,t)$ is independent of time and
thus is simplified as $r(k)=(r_x (k),r_y (k),r_z (k))^T$.

This work treated a pulse sequence as a series of subsequences.
As shown in Fig. \ref{fig1}, the subsequences can be classified into three types:
(a) A with-RF subsequence which includes an RF pulse,
(b) a with-ADC subsequence which contains a sampling time using analog to digital converters (ADCs), and
(c) a relaxation subsequence which contains no RF pulses and sampling times.
Subsequences with both an RF pulse and a sampling time simultaneously are
not commonly used and thus are not considered.

When general subsequences are processed, iterative update is required for computing the magnetization.
For a small duration $\Delta t$, the transition of the magnetization vector $M(k,t)$
between $t$ and $t+\Delta t$ can be expressed as
\begin{equation}\label{eq2}
\mathcal{M}(k,t+\Delta t)=U(k,t,\Delta t)\mathcal{M}(k,t)
\end{equation}
where $U(k,t,\Delta t)$ is a $4 \times 4$ matrix representation of the transition given in Appendix A,
and $\mathcal{M}(k,t)=(M_x (k,t),M_y (k,t),M_z (k,t),M_0 (k))^T$ contains $M(k,t)$
and the non-zero element of $M_{static} (k)$.
The magnetic field vector $B(t)$ is assumed to be piecewise constant between $t$ and $t+\Delta t$.
By combining Eq. (\ref{eq2}) at $t=t_0,t_1,\ldots,t_{N_{RF}-1}$,
a combined transition between $t_0$ and $t_{N_{RF}}$ can be expressed as
\begin{equation}\label{eq3}
\mathcal{M}(k,t_{N_{RF}})=U(k,t_{N_{RF}-1},\Delta t_{N_{RF}-1}) \cdots U(k,t_1,\Delta t_1)U(k,t_0,\Delta t_0)\mathcal{M}(k,t_0)
\end{equation}
if all durations $\Delta t_j=t_{j+1}-t_j (j=0,\ldots,N_{RF}-1)$ are sufficiently small.

In the cases of the special subsequences which satisfy $B_1 (k,t)=0$ for $t_0 \le t \le t_1$,
the magnetization at $t$ can be computed using the following analytic solutions
\begin{align}
M_{xy} (k,t) &= M_{xy} (k,t_0) \exp \Bigg(-\frac{t-t_0}{T_2(k)} \Bigg) \exp \Bigg( \int_{t_0}^t -i\gamma B_z(k,t)dt \Bigg) \label{eq4}\text{, and} \\
   M_z (k,t) &= M_0(k)+(M_z(k,t_0)-M_0(k)) \exp \Bigg(-\frac{t-t_0}{T_1(k)} \Bigg) \label{eq5}
\end{align}
if the $k$-th isochromat is static.
These solutions can be applied to with-ADC and relaxation subsequences.
In with-ADC subsequences, k-space data sampled at the time $t$ are represented as
\begin{equation}\label{eq6}
A(p,t)=\sum_{k=1}^K S(k,p,t) M_{xy}(k,t)
\end{equation}
where $S(k,p,t)$ represents the sensitivity of the $p$-th receiver-coil ($p=1,2,\ldots,P$),
and $M_{xy}(k,t)=M_x (k,t)+iM_y (k,t)$ represents the transverse magnetizations.

\subsection{Computational complexities of conventional methods}

In the cases of the with-RF subsequences, naïve implementations of Eq. (\ref{eq3})
have a complexity of $O(N_{RF} K)$. To accelerate computation of these subsequences,
methods based on combined transitions \cite{Taniguchi1,Taniguchi2,Scholand,Takeshima} can be utilized.
These methods split the update process of Eq. (\ref{eq3}) into the preparation and update steps.

In the preparation step, the combined transition
\begin{equation}\label{eq7}
T(k,t_0,t_{N_{RF}})=U(k,t_{N_{RF}-1},\Delta t_{N_{RF}-1}) \cdots U(k,t_1,\Delta t_1)U(k,t_0,\Delta t_0)
\end{equation}
is computed. The preparation step has a complexity of $O(N_{RF} K)$.
This step can be implemented as either precomputing \cite{Taniguchi1,Taniguchi2,Scholand} or dynamic caching \cite{Takeshima}.
Once $T(k,t_0,t_{N_{RF}})$ is computed, the complexity of the update step is reduced to $O(K)$
since only a single matrix multiplication
$\mathcal{M}(k,t_{N_{RF}}) = T(k,t_0,t_{N_{RF}}) \mathcal{M}(k,t_0)$
is required for all $K$ isochromats.
While the complexity of the update step is lower than that of the naïve implementations,
the complexity of the preparation step is not changed.
When isochromats are static, the with-ADC and relaxation subsequences have complexities of
$O(N_{ADC} PK)$ and $O(K)$, respectively. According to the Eq. (\ref{eq4}) and Eq. (\ref{eq5}),
$M(k,t)$ can be computed directly within $t_0 \le t \le t_1$.
In the cases of the with-ADC sequences,
$N_{ADC}$ represents the number of ADC samples to be acquired with Eq. (\ref{eq6}).

The main load of computation is the simulations of with-RF and with-ADC subsequences.
In realistic sequences to be simulated, the typical number of \linebreak%%%
isochromats $K$ is in the order of millions.
The typical numbers of RF update steps $N_{RF}$ and ADC samples $N_{ADC}$ are in the range of dozens to thousands.

\subsection{Efficient simulations of with-RF and with-ADC subsequences}

The computational complexities can be reduced if those of with-RF and with-ADC subsequences are reduced.
This subsection shows that the computational complexities of simulations can be reduced under
the following assumption: when a subsequence is simulated,
there are several groups where multiple isochromats share $U(k,t,\Delta t$).
This assumption means that the underlying components $B_1(k,t)$, $B_z(k,t)$, $T_1(k)$ and $T_2 (k)$ are all shared.

In the cases of with-RF subsequences under this assumption,
the preparation step can be implemented as computing $T(k,t_0,t_{N_{RF}})$ \emph{once in each group}.
The combined transition can be used in the update step of all isochromats in the group.
Therefore, the complexity of the preparation step can be reduced from $O(N_{RF} K)$ to $O(N_{RF} K_{group})$
where $K_{group}$ represents the number of the groups.
The complexity of the update step is still $O(K)$ since the same update step is used.

In the cases of with-ADC subsequences under this assumption, Eq. (\ref{eq6}) can be rewritten as
\begin{equation}\label{eq8}
A(p,t)=\sum_{k_{group}=1}^{K_{group}} A_{block} (k_{group},p,t)
\end{equation}
where $A_{block}(k_{group},p,t)$ is defined as
\begin{equation}\label{eq9}
A_{block}(k_{group},p,t)=\sum_{k \in L(k_{group})} S(k,p,t)M_{xy}(k,t)
\end{equation}
and $L(k_{group})$ represents a set of isochromats in the $k_{group}$-th group.
By assigning Eq. (\ref{eq4}) into Eq. (\ref{eq9}), $A_{block} (k_{group},p,t)$ for $t_0 \le t \le t_1$ can be represented as
\begin{align}
A_{block} (k_{group},p,t)= &\exp \Bigg( - \frac{t-t_0}{T_2 (k_{group})} \Bigg) \cdot \nonumber \\
                           &\exp \Bigg( \int_{t_0}^t -i \gamma B_z (k_{group},t)dt \Bigg) A_{block}(k_{group},p,t_0 ) \label{eq10}
\end{align}
where $T_2 (k_{group})$ and $B_z (k_{group},t)$ represent shared components in the $k_{group}$-th group.

By utilizing Eq. (\ref{eq10}), with-ADC subsequences can be simulated using the \linebreak%%%
following three-step method.
As the first step whose time is $t_0$, compute \linebreak%%%
$A_{block} (k_{group}, p, t_0)$ using Eq. (\ref{eq9}).
This step reduces $K$ isochromats into $P K_{group}$ groups.
As the second step, compute $N_{ADC}$ sampling values using Eq. (\ref{eq8}) and Eq. (\ref{eq10}) for $P K_{group}$ groups.
As the third step, update $M(k,t)$ using Eq. (\ref{eq4}) and Eq. (\ref{eq5}) for $K$ isochromats.
The first, second and third steps have complexities of $O(PK)$, $O(N_{ADC} P K_{group})$ and $O(K)$, respectively.
\begin{figure}[t]%
\centering%
\includegraphics[width=8cm]{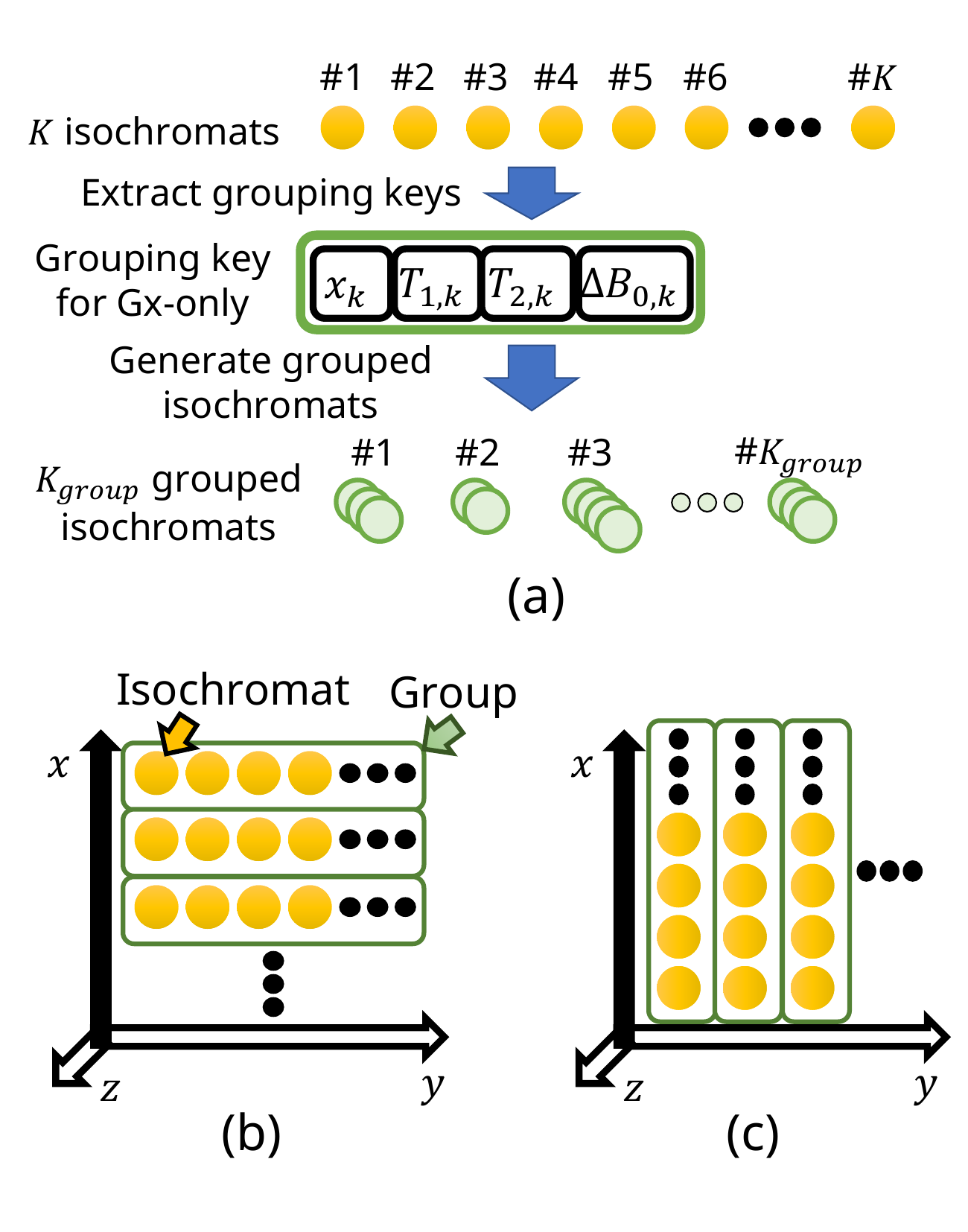}
\caption{Generation of grouped isochromats.
(a) A process flow for generating grouped isochromats.
For each isochromat, values which should be shared are extracted as the grouping key.
The isochromats whose grouping keys are same are grouped.
(b) An example case for updating magnetizations efficiently in the cases of Gx-only.
The isochromats in a group shares their $r_x (k,t)$, $T_1 (k)$, $T_2 (k)$ and $\Delta B_0 (k)$ values.
(c) Another example case for updating magnetizations efficiently in the cases of Gy-only.
The isochromats in a group shares their $r_y (k,t)$, $T_1 (k)$, $T_2 (k)$ and $\Delta B_0 (k)$ values.}\label{fig2}
\end{figure}
\subsection{Grouping criteria for efficient simulations}%

This subsection shows simple criteria for grouping isochromats.
It is assumed that the grouping process of the isochromats are performed once.
Unless the computational times of the grouping operation is too great,
the simulations can be accelerated if types of subsequences can be identified with trivial computational times.

This work considers special gradient types found in common pulse sequences: Gx-only, Gy-only, Gz-only, and no-grads.
These criteria assume that with-RF and with-ADC subsequences tend to have only one active component of G(t).
As shown in Fig. \ref{fig1}, when the gradient field $G_x (t)$ is non-zero,
the remaining gradient fields $G_y (t)$ and $G_z (t)$ are sometimes zeros in the subsequences to be simulated.
Under this gradient type, isochromats can be grouped
when $r_x (k,t)$, $T_1(k)$, $T_2(k)$ and $\Delta B_0(k)$ values are the same values,
as shown in Fig. \ref{fig2}. This gradient type is referred to as Gx-only.
Similar gradient types with single non-zero gradient fields of $G_y (t)$ and $G_z (t)$ can be defined
in the cases of Gy-only and Gz-only, respectively.
The no-grads type represents subsequences whose gradient fields are zeros.
Its example cases are chemical shift selective (CHESS) pulses \cite{CHESS}
and sequences for adjusting RF power levels \cite{Perman,Carlson}.

The difference between the proposed and conventional methods
is the strategy for updating multiple isochromats simultaneously.
The conventional methods accelerated simulations by updating them
in parallel with various hardware \cite{SIMRI,JEMRIS,MRISIMUL,MRiLab,Kose,coreMRI,KomaMRI,Takeshima}.
In contrast, the proposed method updates them by utilizing grouped isochromats
and thus is applicable to both non-parallel and parallel computing environments.

\section{Methods}

To evaluate the performance of MR simulation methods,
the proposed method was implemented as a part of house-made software \cite{Takeshima} named a virtual MR scanner (VMRscan).
The software has a conventional acceleration method using combined transitions
with a dynamic caching method for simulating with-RF subsequences.
For utilizing grouped isochromats, the proposed method used three different flows.
The software was run on a central processing unit (CPU) with 8 performance cores,
16 efficient cores and 32 processor threads.
The frequencies of the CPU were 3.2 GHz for the performance cores and 2.4 GHz for the efficient cores.
These cores were dynamically boosted up to 6.0 GHz.

In addition to the simulation method, evaluations require phantoms and sequences.
As phantoms, numeric and measured phantoms were developed.
The numeric phantom was the one used in the previous work \cite{Takeshima}.
The measured phantom was generated with a simple parameter mapping method and
preprocessed with a clustering algorithm.
For the sake of evaluations, spin-echo (SE), fast SE (FSE) and echo-planar imaging (EPI)
were developed in the Pulseq format \cite{Pulseq}.

The efficiency was evaluated using the VMRscan with and without the proposed method.
The relationship between the number of isochromats and the computational time was also evaluated.

\subsection{Fast simulations with grouped isochromats}%
\begin{figure}[t]%
\centering%
\includegraphics[width=\linewidth]{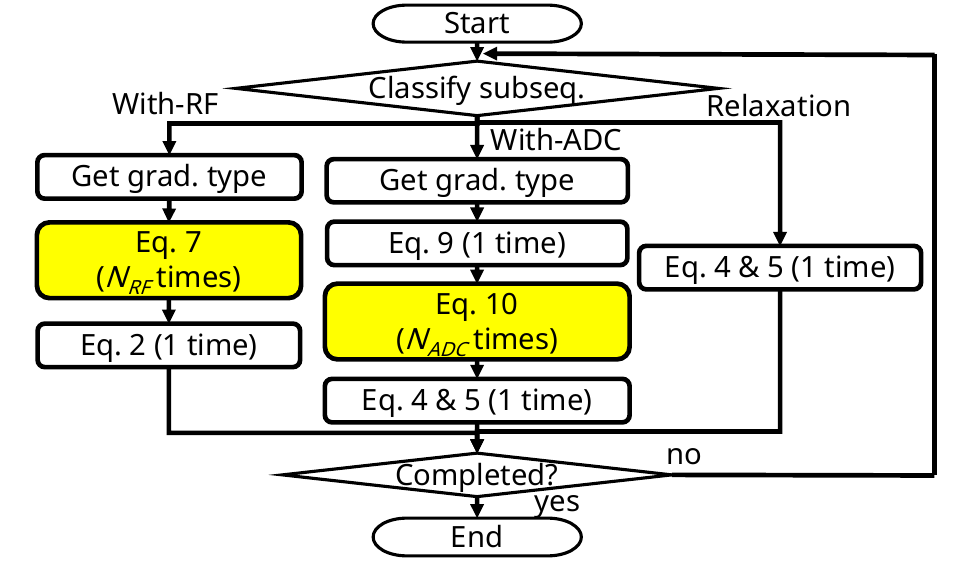}
\caption{The overview of the proposed method.
Each subsequence is classified into three types.
In the cases of with-RF and with-ADC subsequences,
grouped isochromats are used in time-consuming processes.
When the grouping criteria are not met in a subsequence,
the subsequence is simulated using conventional methods (not shown in this figure).
Abbreviations: RF, radiofrequency, ADC, analog-to-digital converter.}\label{fig3}
\end{figure}

The proposed method split a sequence into a series of subsequences,
classified the subsequence individually,
and simulates the subsequences using one of three different flows as shown in Fig. \ref{fig3}.
In the cases of with-RF and with-ADC subsequences,
gradient types of the subsequences were classified before simulating the subsequences.
Whenever gradient types of subsequences were classified as one of known types,
grouped simulations explained in the theory were utilized.
Remaining subsequences were simulated with a conventional method \cite{Takeshima}.

The relaxation subsequences could be processed quickly
regardless of whether these subsequences had specific gradient types or not.
The accelerations using specific gradient types were applied to the with-RF and with-ADC subsequences only.
It is worth noting that practical pulse sequences often used overlapped gradients for reducing TE
as shown in Fig. \ref{fig1}.
Such overlapped gradients don't affect the performance of the proposed method which accelerates
with-RF and with-ADC subsequences since the overlapped gradients are commonly used in relaxation subsequences.

\subsection{Numeric and measured phantoms}%
\begin{table}[t]%
\caption{Pulse sequences used in acquisitions of a volunteer.
The acquired images were used for creating the brain phantom.
Abbreviations: T1W, T1-weighted, T2W, T2-weighted, SE, spin echo, FSE, fast spin echo.}\label{table1}
\centering%
\begin{tabular}{|l|l|l|} \hline
 & T1W (4 acquisitions) & T2W \\ \hline
Sequence type & SE & FSE with 4 echos \\ \hline
Reconstruction matrix & $512 \times 480 \times 48$ & $512 \times 480 \times 48$ \\ \hline
FOV & $256 \times 240 \times 240 \,\text{mm}^3$ & $256 \times 240 \times 240 \,\text{mm}^3$ \\ \hline
Num. echos & 1 & 4 \\ \hline
TE & 10 ms & 20, 60, 100, 140 ms \\ \hline
TR & 500, 800, 1100, 1500 ms & 4500 ms \\ \hline
\end{tabular}
\end{table}

A numeric phantom named circles was used as a phantom suitable for the proposed method.
The phantom contained a large circle and 9 small circles.
These small circles were put in the large circle.
The matrix and spatial sizes of the phantom were $512 \times 512 \times 10$
and $256 \,\text{mm} \times 256 \,\text{mm} \times 10 \,\text{mm}$, respectively.
In the phantom, 1189180 isochromats (1 isochromat per pixel) were put within the large circle.
In each circle, $T_1 (k)$, $T_2 (k)$ and $\Delta B_0 (k)$ values were constants.
Number of groups were 10, 1128, 1124 and 100 for no-grads, Gx-only, Gy-only, and Gz-only, respectively.
This phantom is suitable for simulating sequences with small numbers of groups in all gradient types.

A measured phantom named brain was generated from 8 multi-slice images of a brain.
These images were acquired in a volunteer scan.
The pulse sequences used in the scan are shown in Table \ref{table1}.
These images consisted of four T1-weighted (T1W) images, and four T2-weighted (T2W) images.
All acquisitions shared the matrix and field-of-view (FOV) sizes.
The number of readouts, phase encodes, and slices were 256, 160 and 48, respectively.
The size of reconstructed images was $512 \times 480 \times 48$ by zero-filling unknown k-space data
for scaling the images in the readout and phase encoding directions.
The FOV size was $256 \,\text{mm} \times 240 \,\text{mm} \times 240 \,\text{mm}$.
The $T_1 (k)$ and $T_2 (k)$ values were estimated using the following parameter mapping method.
For each pixel in the $T_1$ and $T_2$ maps, the parameter mapping method
searched the best value from a set of discrete candidate values.
The candidate values were put in the range of 10 to 3000 milliseconds
for the $T_1$ map, and 10 to 500 milliseconds for the $T_2$ map.
For individual maps, 100 logarithmically spaced candidate values were used.
The $M_0 (k)$ values were estimated using the T2W images, $T_1$ map and $T_2$ map.
Any voxels whose $M_0(k)$ values were lower than a threshold were ignored.
The $\Delta B_0 (k)$ values were set to 0 Hz.
The volunteer scan was approved by our institutional review board
and informed consent was obtained from the volunteer.

\subsection{Preprocessing method for grouping isochromats efficiently}%
\begin{figure}[t]%
\centering%
\includegraphics[width=\linewidth]{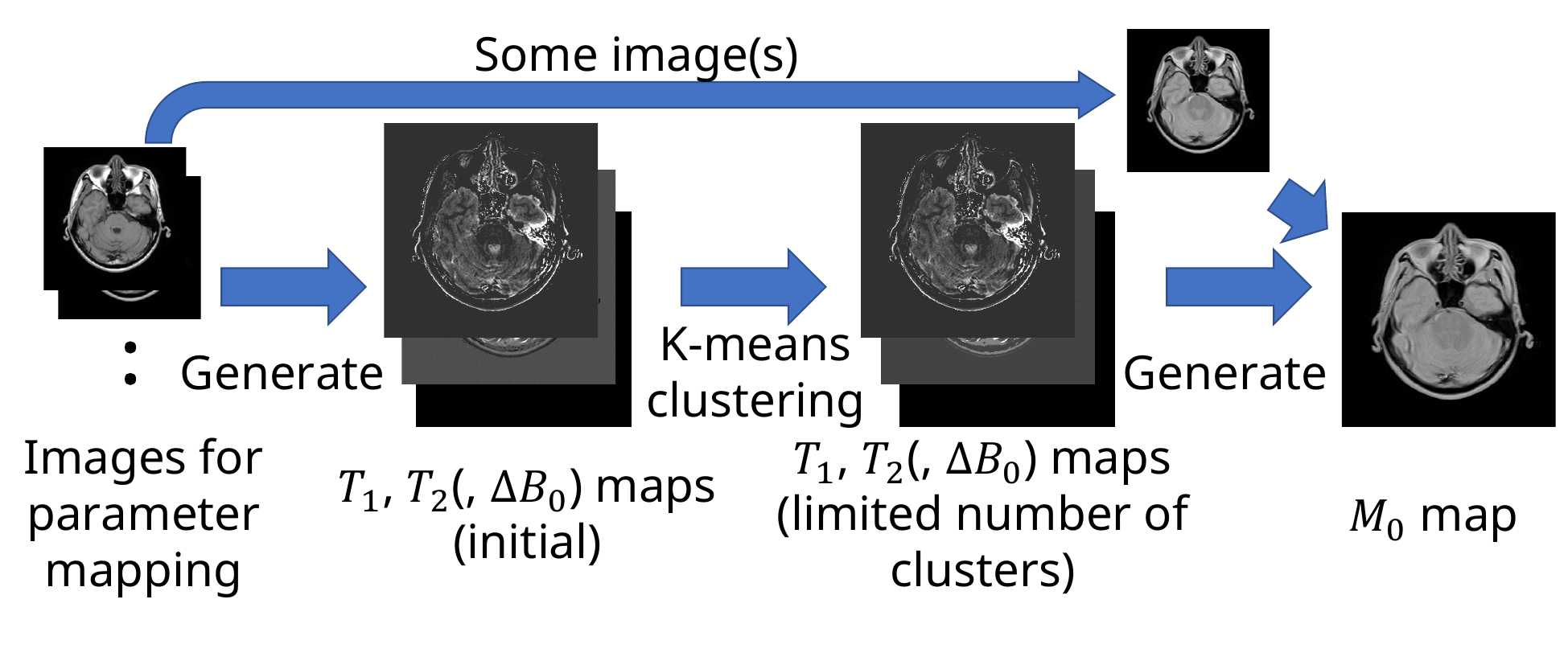}%
\caption{Preprocessing method for ensuring that the numbers of groups are sufficiently small.
For reducing the number of groups,
a clustering algorithm was applied to the initial maps consisting of $T_1$ and $T_2$ maps.
The $\Delta B_0$ map was not used and was zero-filled in the evaluations.
The remaining map $M_0$ was generated using the clustered parameters
and existing images used in the parameter mapping of the initial parameter maps.}\label{fig4}
\end{figure}

When the number of groups is great, the proposed method cannot reduce computational time of simulations.
In the cases of numerical phantoms such as the circles phantom,
all numbers of groups using above-mentioned criteria can be controlled by design.
However, in general cases, the numbers of groups are close to the number of isochromats.
To use the proposed method efficiently, phantoms could be preprocessed
if it is acceptable to change time constants and field inhomogeneity of isochromats slightly.

To ensure that the numbers of groups are sufficiently small,
phantoms can be preprocessed using an arbitrary clustering algorithm.
The overview of the preprocessing method is shown in Fig. \ref{fig4}.
Initial parameter maps can be generated by an arbitrary parameter mapping method.
To generate parameter maps with a limited number of parameters,
a clustering algorithm was applied to the initial parameter maps.
The space of the clustering algorithm was $(T_1,T_2,\Delta B_0)$.
As the clustering algorithm, a bisecting K-means clustering algorithm
in the scikit-learn package \cite{ScikitLearn} was used.
The $M_0(k)$ values were refined using the T2W images and the center values of the updated parameter maps.
The locations $r(k)$ were also aligned when the isochromats were put into voxels.

\subsection{Pulse sequences}
\begin{table}[t]%
\caption{Pulse sequences used in simulations.
Four different sequences were used for evaluations.
These sequences were written in the Pulseq format.
Abbreviations: T1W, T1-weighted, T2W, T2-weighted,
EPI, echo planar imaging, FSE, fast spin echo, ETL, echo train length.}\label{table2}
\centering%
\begin{tabular}{|l|r|r|r|r|} \hline
 & T1W & T2W & EPI & FSE \\ \hline
Sequence type & \multicolumn{2}{|c|}{Spin echo} & Spin echo EPI & Fast spin echo \\ \hline
RF pulse type & \multicolumn{4}{|c|}{Hamming-windowed sinc} \\ \hline
RF pulse (excitation) & \multicolumn{4}{|c|}{4000 us, 90 degrees, 1 us/sample} \\ \hline
RF pulse (refocusing) & \multicolumn{4}{|c|}{4000 us, 180 degrees, 1 us/sample} \\ \hline
Num. with-RF & 512 & 512 & 2 & 312 \\ \hline
Num. with-ADC & 256 & 256 & 96 & 260 \\ \hline
FOV & \multicolumn{4}{|c|}{256 mm x 256 mm} \\ \hline
Slice thickness & \multicolumn{4}{|c|}{10 mm} \\ \hline
TR & 500 ms & 4500 ms & N/A & 1500 ms \\ \hline
TE & 10 ms & 100 ms & 70 ms & 36 ms \\ \hline
Num. readout samples & 256 & 256 & 128 & 256 \\ \hline
%Num. phase encodes & 256 & 256 & 96 (partial Fourier) & 260 (52 TRs, ETL 5) \\ \hline
\multirow{2}{*}{Num. phase encodes} & \multirow{2}{*}{256} & \multirow{2}{*}{256} & 96 (partial & 260 (52 TRs, \\
 &  &  & Fourier) & ETL 5) \\ \hline
Num. slices & \multicolumn{4}{|c|}{1} \\ \hline
Readout sampling  & \multirow{2}{*}{5} & \multirow{2}{*}{50} & \multirow{2}{*}{3} & \multirow{2}{*}{8} \\
ratio (us/sample) &   &    &   &   \\ \hline
\end{tabular}
\end{table}

To evaluate the computational times depending on the type of subsequences,
SE, FSE and EPI sequences were developed.
The SE sequence was used for simulating acquisitions with T1W and T2W contrasts.
The parameters used in these sequences are given in Table \ref{table2}.
The SE sequence acquired 256 TRs. It consisted of 512 with-RF and 256 with-ADC subsequences.
The FSE sequence acquired 52 TRs with echo train length (ETL) of 5.
It consisted of 312 with-RF and 260 with-ADC subsequences.
The EPI sequence consisted of 2 with-RF and 96 with-ADC subsequences.
The SE and FSE sequences were implemented as full-sampling sequences.
The numbers of readouts and phase encodes were both 256.
The EPI sequence was implemented as a single-shot sequence with a partial-Fourier acquisition.
The number of readouts was 128. The k-space size was 128 in the phase encode direction.
With the partial-Fourier acquisition, the number of acquisitions was reduced to 96.

\subsection{Simulations \#1}

To evaluate the efficiency of the proposed method,
processing times of the following simulations were measured 6 times.
In the cases of the circle phantom, simulations with and without the proposed method were evaluated.
In the cases of the brain phantom, simulations with the proposed method
were performed using the preprocessing method with 64, 128 and 256 clusters.
These phantoms are referred to as the brain-64, brain-128 and brain-256 phantoms, respectively.
Simulations without the proposed method used the brain phantom without preprocessing methods.
All simulations were evaluated with and without single instruction multiple data (SIMD) instructions.
In all the simulations, multi-threading implemented in the previous work were used \cite{Takeshima}.
The k-space data acquired with the T1W, T2W and FSE sequences were reconstructed using a fast Fourier transform.
The k-space data acquired with the EPI sequence were reconstructed
using a gridding method \cite{Jackson}.

\subsection{Simulations \#2}

To evaluate the relationship between the number of isochromats and the computational time,
processing times using the brain phantoms were also measured with the following subvoxel types:
$1 \times 1 \times 1$,
$2 \times 1 \times 1$,
$2 \times 2 \times 1$,
$2 \times 2 \times 2$.
These numbers represent the numbers of divided subvoxels in x, y, and z directions.
All simulations were evaluated with SIMD instructions.
These times were measurements once for each condition.
Other conditions were same as those in the evaluations \#1.

\section{Results}

\subsection{Simulations \#1}%
\begin{table}[t]%
\caption{Average processing times for simulating T1W, T2W, EPI and FSE sequences
in the cases of (a) circles and (b) brain.
Each ratio of the processing time of the proposed method
to that of the conventional method is shown in parentheses.
The simulation times of the proposed method were 4 to 72 times
faster than those of the conventional methods.
Abbreviations: T1W, T1-weighted, T2W, T2-weighted,
EPI, echo planar imaging, FSE, fast spin echo, SIMD, single instruction multiple data.}\label{table3}
\centering%
\begin{tabular}{|l|r|r|r|r|} 
\multicolumn{5}{l}{(a) Circles} \\ \hline
 & T1W & T2W & EPI & FSE \\ \hline
Ungrouped & 55.04 & 68.04 & 17.99 & 58.09 \\ \hline
10 groups & 2.72 (20.21) & 14.37 (4.74) & 0.25 (72.05) & 2.54 (22.86) \\ \hline
{\small Ungrouped, SIMD} & 10.10 & 10.13 & 3.09 & 10.55 \\ \hline
{\small 10 groups, SIMD} & 1.41 (7.15) & 1.44 (7.05) & 0.15 (20.75) & 1.32 (8.00) \\ \hline
\multicolumn{5}{l}{(b) Brain} \\ \hline
 & T1W & T2W & EPI & FSE \\ \hline
Ungrouped & 155.75 & 160.41 & 52.17 & 164.74 \\ \hline
64 groups & 7.43 (20.96) & 18.71 (8.58) & 0.86 (60.86) & 6.90 (23.86) \\ \hline
128 groups & 8.44 (18.45) & 16.67 (9.62) & 1.10 (47.41) & 8.00 (20.59) \\ \hline
256 groups & 10.31 (15.10) & 17.97 (8.93) & 1.60 (32.51) & 9.86 (16.71) \\ \hline
{\small Ungrouped, SIMD} & 28.51 & 29.07 & 8.88 & 29.35 \\ \hline
{\small 64 groups, SIMD} & 5.24 (5.44) & 7.60 (3.83) & 0.66 (13.37) & 4.55 (6.45) \\ \hline
{\small 128 groups, SIMD} & 5.57 (5.12) & 6.47 (4.49) & 0.77 (11.47) & 4.94 (5.93) \\ \hline
{\small 256 groups, SIMD} & 6.18 (4.61) & 7.69 (3.78) & 1.01 (8.82) & 5.60 (5.24) \\ \hline
\end{tabular}
\end{table}%
\begin{table}[t]%
\caption{Numbers of isochromats and groups depending on subvoxel types.
In the cases of no-grads, the numbers of groups were constants.
In the other cases, numbers of groups were approximately proportional to the number of subvoxels per voxel.}\label{table4}
\centering%
\begin{tabular}{|l|r|r|r|r|} \hline
Subvoxel Type & $1 \times 1 \times 1$ & $2 \times 1 \times 1$ & $2 \times 2 \times 1$ & $2 \times 2 \times 2$ \\ \hline
Num. isochromats & 3,495,411  & 6,943,821  & 13,887,763  & 27,541,079 \\ \hline
No-grads (brain-64) & 64  & 64  & 64  & 64 \\ \hline
Gx-only (brain-64) & 21,046  & 42,762  & 43,076  & 43,891 \\ \hline
Gy-only (brain-64) & 26,296  & 27,126  & 54,028  & 52,734 \\ \hline
Gz-only (brain-64) & 2,715  & 2,755  & 2,757  & 5,575 \\ \hline
No-grads (brain-128) & 128  & 128  & 128  & 128 \\ \hline
Gx-only (brain-128) & 41,384  & 82,766  & 84,289  & 85,333 \\ \hline
Gy-only (brain-128) & 50,867  & 51,993  & 103,540  & 101,466 \\ \hline
Gz-only (brain-128) & 5,406  & 5,427  & 5,443  & 10,998 \\ \hline
No-grads (brain-256) & 256  & 256  & 256  & 256 \\ \hline
Gx-only (brain-256) & 78,221  & 157,777  & 162,980  & 166,452 \\ \hline
Gy-only (brain-256) & 94,417  & 98,777  & 196,048  & 195,764 \\ \hline
Gz-only (brain-256) & 10,580  & 10,715  & 10,764  & 21,810 \\ \hline
\end{tabular}
\end{table}%
\begin{table}[tp]%
\caption{Average processing times when a voxel was divided into subvoxels.
Each ratio of the processing time of the proposed method to
that of the conventional method is shown in parentheses.
Abbreviations: T1W, T1-weighted, T2W, T2-weighted, EPI, echo planar imaging, FSE, fast spin echo.}\label{table5}
\centering%
\begin{tabular}{|l|r|r|r|r|}
\multicolumn{5}{l}{(a) Processing times in second (T1W)} \\ \hline
Subvoxel Type & 1x1x1 & 2x1x1 & 2x2x1 & 2x2x2 \\ \hline
Brain (ungrouped) & 28.61 & 55.07 & 106.59 & 204.83 \\ \hline
Brain-64 & 5.24 (5.46) & 10.70 (5.15) & 20.52 (5.19) & 39.77 (5.15) \\ \hline
Brain-128 & 5.59 (5.12) & 11.55 (4.77) & 21.66 (4.92) & 41.26 (4.96) \\ \hline
Brain-256 & 6.19 (4.62) & 13.21 (4.17) & 23.71 (4.50) & 44.51 (4.60) \\ \hline
\multicolumn{5}{l}{(b) Processing times in second (T2W)} \\ \hline
Subvoxel Type & 1x1x1 & 2x1x1 & 2x2x1 & 2x2x2 \\ \hline
Brain (ungrouped) & 29.08 & 55.56 & 108.04 & 206.61 \\ \hline
Brain-64 & 7.60 (3.83) & 12.08 (4.60) & 24.78 (4.36) & 51.15 (4.04) \\ \hline
Brain-128 & 6.45 (4.51) & 13.62 (4.08) & 25.65 (4.21) & 51.40 (4.02) \\ \hline
Brain-256 & 7.67 (3.79) & 16.18 (3.43) & 27.98 (3.86) & 54.49 (3.79) \\ \hline
\multicolumn{5}{l}{(c) Processing times in second (EPI)} \\ \hline
Subvoxel Type & 1x1x1 & 2x1x1 & 2x2x1 & 2x2x2 \\ \hline
Brain (ungrouped) & 8.80 & 17.56 & 34.26 & 66.38 \\ \hline
Brain-64 & 0.65 (13.45) & 1.33 (13.25) & 2.43 (14.12) & 4.64 (14.32) \\ \hline
Brain-128 & 0.77 (11.42) & 1.67 (10.52) & 2.97 (11.53) & 5.44 (12.21) \\ \hline
Brain-256 & 1.00 (8.81) & 2.32 (7.56) & 3.85 (8.90) & 7.11 (9.34) \\ \hline
\multicolumn{5}{l}{(d) Processing times in second (FSE)} \\ \hline
Subvoxel Type & 1x1x1 & 2x1x1 & 2x2x1 & 2x2x2 \\ \hline
Brain (ungrouped) & 29.42 & 56.80 & 109.46 & 208.44 \\ \hline
Brain-64 & 4.56 (6.46) & 9.16 (6.20) & 17.19 (6.37) & 32.88 (6.34) \\ \hline
Brain-128 & 4.97 (5.92) & 9.98 (5.69) & 18.30 (5.98) & 35.02 (5.95) \\ \hline
Brain-256 & 5.61 (5.24) & 11.69 (4.86) & 20.42 (5.36) & 38.11 (5.47) \\ \hline
\end{tabular}
\end{table}%
\begin{figure}[t]%
\centering%
\includegraphics[width=8cm]{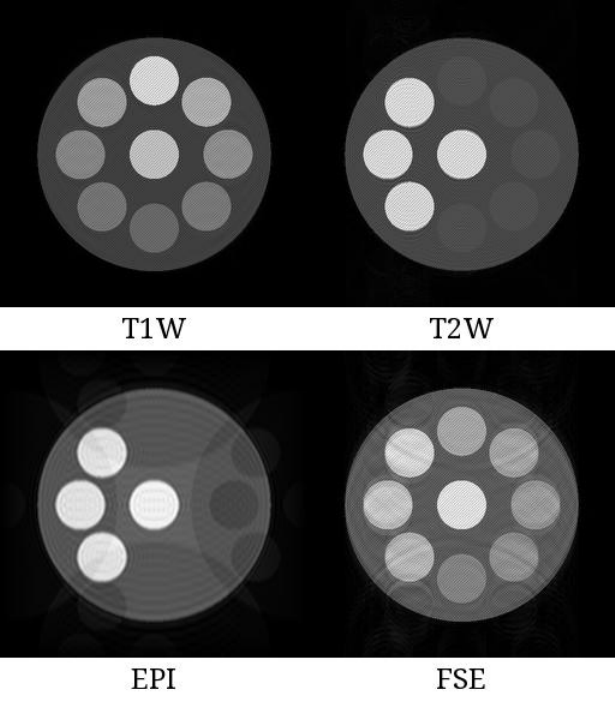}
\caption{Reconstructed images using the circles phantom.
T1W, T2W, EPI and FSE sequences were used for the simulations.
Each circle has a unique pair of $T_1$ and $T_2$ values.
The images using EPI and FSE sequences included artifacts
since no artifact elimination methods were used.
Abbreviations: T1W, T1-weighted, T2W, T2-weighted, EPI, echo-planar imaging, FSE, fast spin echo.}\label{fig5}
\end{figure}
\begin{figure}[t]%
\centering%
\includegraphics[width=\linewidth]{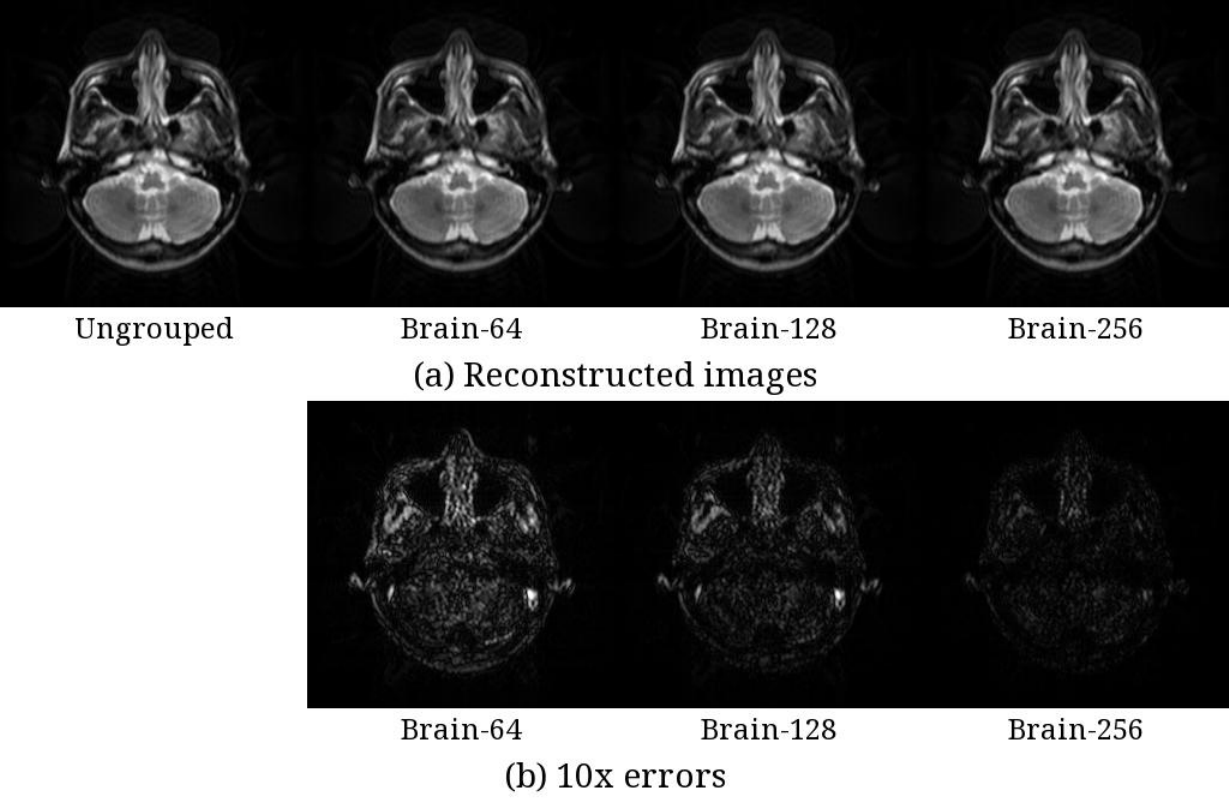}
\caption{Reconstructed images and their error images in the case of an EPI sequence.
(a) Reconstructed images using the brain phantom with reference (ungrouped),
64 clusters, 128 clusters and 256 clusters.
(b) Errors between the reference and grouped images.
The errors were multiplied by 10. As shown in the error images,
the image with 256 clusters were visually close to the reference image.
Abbreviations: EPI, echo-planar imaging.}\label{fig6}
\end{figure}
\begin{figure}[t]%
\centering%
\includegraphics[width=\linewidth]{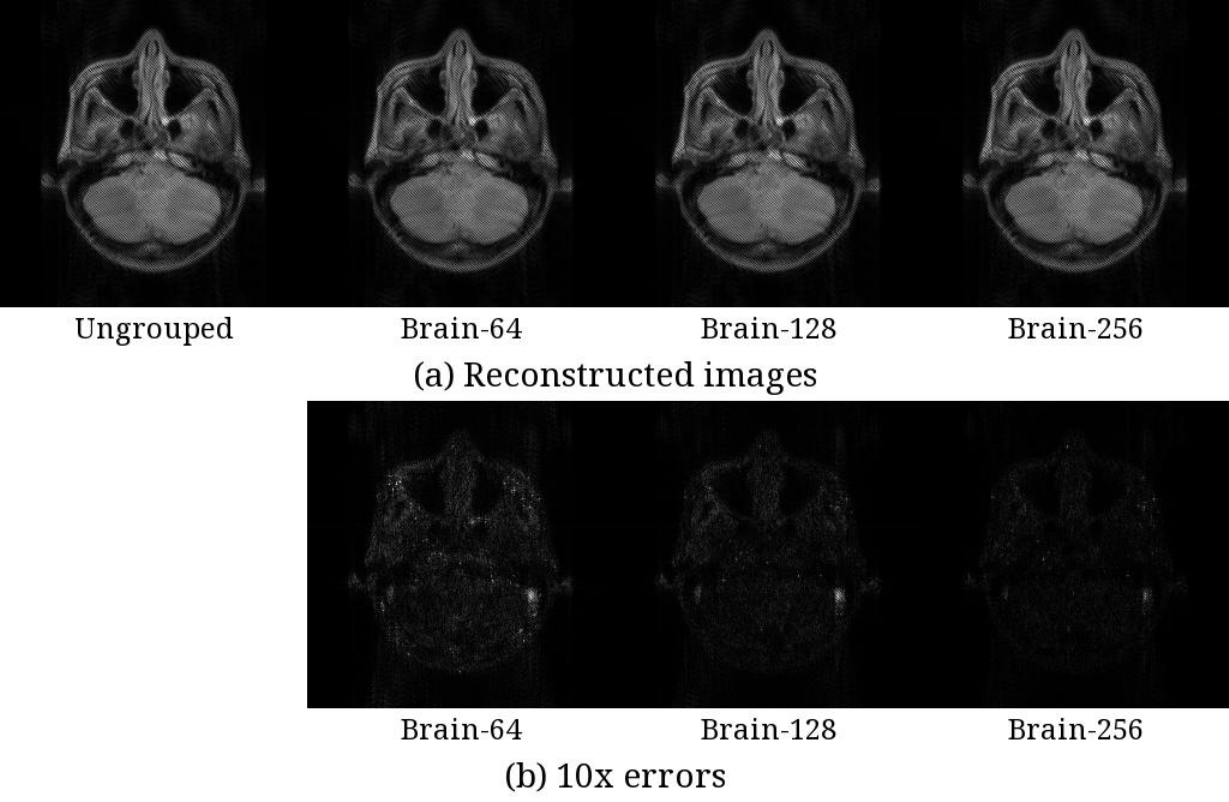}
\caption{Reconstructed images and their error images in the case of an FSE sequence.
(a) Reconstructed images using the brain phantom with reference (ungrouped),
64 clusters, 128 clusters and 256 clusters.
(b) Errors between the reference and grouped images.
The errors were multiplied by 10.
As shown in the error images,
all image with 64 and more clusters were visually close to the reference image.
Abbreviations: FSE, fast spin echo.}\label{fig7}
\end{figure}

The average processing times are shown in Table \ref{table3}.
Number of isochromats and groups used in the brain phantom are shown
in the $1 \times 1 \times 1$ column of Table \ref{table4}.
Reconstructed images using the circles phantom are shown in Fig. \ref{fig5}.
In the cases of the brain phantoms, reconstructed images and their error images are
shown in Fig. \ref{fig6}, Fig. \ref{fig7}, Fig. S1 and Fig. S2
for EPI, FSE, T1W and T2W sequences, respectively.
In the cases of the brain phantoms with 3.5 million isochromats,
the conventional method \cite{Takeshima} with SIMD instructions simulated
T1W, T2W, EPI and FSE sequences in 28.5, 29.1, 8.9 and 29.4 seconds, respectively.
In the cases of 256 clusters, the proposed method simulated
these sequences in 8.2, 7.7, 1.0 and 5.6 seconds, respectively.
The simulation times of the proposed method were 4 to 72 times faster than those of the conventional methods.

As shown in the reconstruction error images,
the image with 256 clusters were visually close to the image without clustering (reference image).
The errors were high when sequences with long TR values (T2W and EPI sequences) were simulated.
The errors were decreased when the number of clusters was increased.

\subsection{Simulations \#2}

The processing times are shown in Table \ref{table5}.
Number of isochromats and groups used in these simulations are shown in Table \ref{table4}.
Reconstructed images are shown in supplementary materials Fig. S3-S6.
The simulation times of the proposed method were 3 to 14 times faster than those of the conventional methods.
In the cases of 27.5 million isochromats using single instruction multiple data (SIMD) instructions and multi-threading,
the conventional method simulated FSE and EPI sequences in 208.4 and 66.4 seconds, respectively.
In the same cases, the proposed method simulated these sequences in 38.1 and 7.1 seconds, respectively.
In all evaluated cases, the processing times were approximately proportional to the number of isochromats.

The numbers of groups with single non-zero gradient fields were in the range of 2.7 to 196 thousand.
When the number of subvoxels was doubled in a gradient axis,
the number of groups was approximately doubled in the axis.

\section{Discussion}

Both evaluations showed that the proposed method further accelerated a conventional acceleration method
using combined transitions \cite{Takeshima}.
As shown in Table \ref{table3} and Table \ref{table5},
the proposed method accelerated simulations both with and without SIMD instructions
in the cases of both the SE and EPI sequences.
These results showed that the proposed method efficiently reduced the processing times
which included the complexities of $O(N_{RF} K)$ and $O(N_{ADC} P K)$.

The proposed method enlarges suitable environments for simulations.
The results of the evaluation \#1 also showed that
the proposed method could simulate all tested sequences efficiently without SIMD instructions.
The proposed method reduced the gaps between the processing times with and without special hardware.

The proposed method has a capability of accelerating sequences without using RF pulses repeatedly.
Existing methods \cite{Taniguchi1,Taniguchi2,Scholand,Takeshima}
couldn't accelerate such sequences
since these methods accelerated simulations by reusing combined transitions with later subsequences.
Both Table \ref{table3} and Table \ref{table5} showed that the proposed method was efficient to accelerate the EPI sequence
by sharing combined transitions in multiple isochromats.

The major source of the computational complexity was still the number of isochromats for the proposed method.
While the increases in the number of clusters affected the processing times,
their impacts were small as shown in the results of the evaluation \#2.

There is a trade-off between reduction of processing times and loss of faithfulness of measured phantoms.
The results showed that errors were decreased gradually when the number of clusters were increased.
Since the MR simulators cannot control sequences to be simulated,
it is preferred for avoiding the loss of the faithfulness to use moderately large numbers of clusters
such as 256 clusters.
This trade-off is not a matter in the cases of numeric phantoms.

The remained work is to accelerate subsequences which cannot be classified as current gradient types.
In such subsequences,
the number of groups is much increased since two or more components in the 3-dimensional space must be shared.
While it is easy to use additional gradient types,
it is not easy to accelerate simulations efficiently with such groups.

\section{Conclusions}

A new computation method using grouped isochromats for accelerating the simulations has been proposed.
The proposed method using grouped isochromats was efficient
for reducing the computational times of the simulations.
Experimental results showed that the simulation times of the proposed method were
3 to 72 times faster than those of the conventional methods.

\section{CRediT authorship contribution statement}

Hidenori Takeshima: Conceptualization, Data curation, Formal analysis, Investigation, Methodology, Software, Validation, Writing – original draft

\section{Declaration of competing interest}

The author declares the following financial interests/personal relationships \linebreak%%%
which may be considered as potential competing interests:
Hidenori Takeshima reports financial support was provided by Canon Medical Systems Corporation.
Hidenori Takeshima has patent pending to Canon Medical Systems Corporation.

\appendix
\section{Matrix representation of transition}

In this paper, the implementations used an asymmetric operator splitting \linebreak%%%
method \cite{Graf2}.
In this method, the transition $U(k,t,\Delta t)$ is given as
$U(k,t,\Delta t) = $ \linebreak%%%
$D(k, \Delta t) R(k, t, t+\Delta t)$.
The matrices used in $U(k,t,\Delta t)$ are given as
\begin{equation}\label{eqA1}
D=
\begin{pmatrix}
\exp(-\Delta t \text{⁄} T_2) & 0 & 0 & 0 \\
0 & \exp(-\Delta t \text{⁄} T_2) & 0 & 0 \\
0 & 0 & \exp(-\Delta t \text{⁄} T_1) & 1 - \exp(-\Delta t \text{⁄} T_1) \\
0 & 0 & 0 &1
\end{pmatrix}
\text{, and}
\end{equation}
\footnotesize
\begin{equation}\label{eqA2}
R=
\begin{pmatrix}
      \cos \theta + u_x u_x(1 - \cos \theta) &
- u_z \sin \theta + u_x u_y(1 - \cos \theta) &
  u_y \sin \theta + u_x u_z(1 - \cos \theta) &
  0 \\
  u_z \sin \theta + u_x u_y(1 - \cos \theta) &
      \cos \theta + u_y u_y (1 - \cos \theta) &
- u_x \sin \theta + u_y u_z (1 - \cos \theta) & 
  0 \\
- u_y \sin \theta + u_x u_z (1 - \cos \theta) &
  u_x \sin \theta + u_y u_z (1 - \cos \theta) &
      \cos \theta + u_z u_z (1 - \cos \theta) &
  0 \\
0 & 0 & 0 & 1
\end{pmatrix}
\end{equation}
\normalsize

where $(u_x,u_y,u_z) = B(k,t) \text{⁄} |B(k,t)|$ and $ \theta = -\gamma |B(k,t)|\Delta t$.

\bibliographystyle{elsarticle-num}
\bibliography{paperref}

\newpage
\section{Supplementary Materials}

\begin{figure}[ht]%
{\centering%
\includegraphics[width=\linewidth]{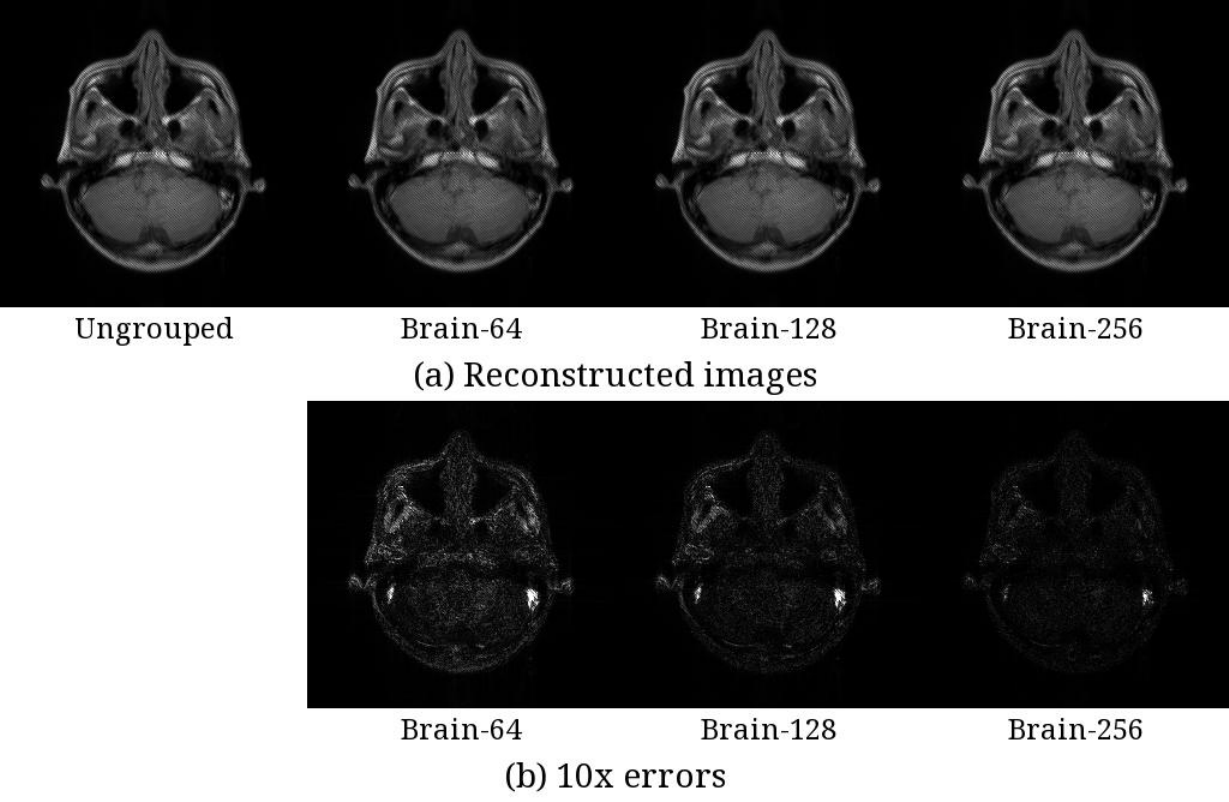}
}%
Figure S1.
Reconstructed images and their error images in the case of a T1W sequence.
(a) Reconstructed images using the brain phantom with reference (ungrouped),
64 clusters, 128 clusters and 256 clusters.
(b) Errors between the reference and grouped images.
The errors were multiplied by 10.
As shown in the error images, all image with 128 and more clusters were
visually close to the reference image. Abbreviations: T1W, T1-weighted.
\end{figure}

\begin{figure}[ht]%
{\centering%
\includegraphics[width=\linewidth]{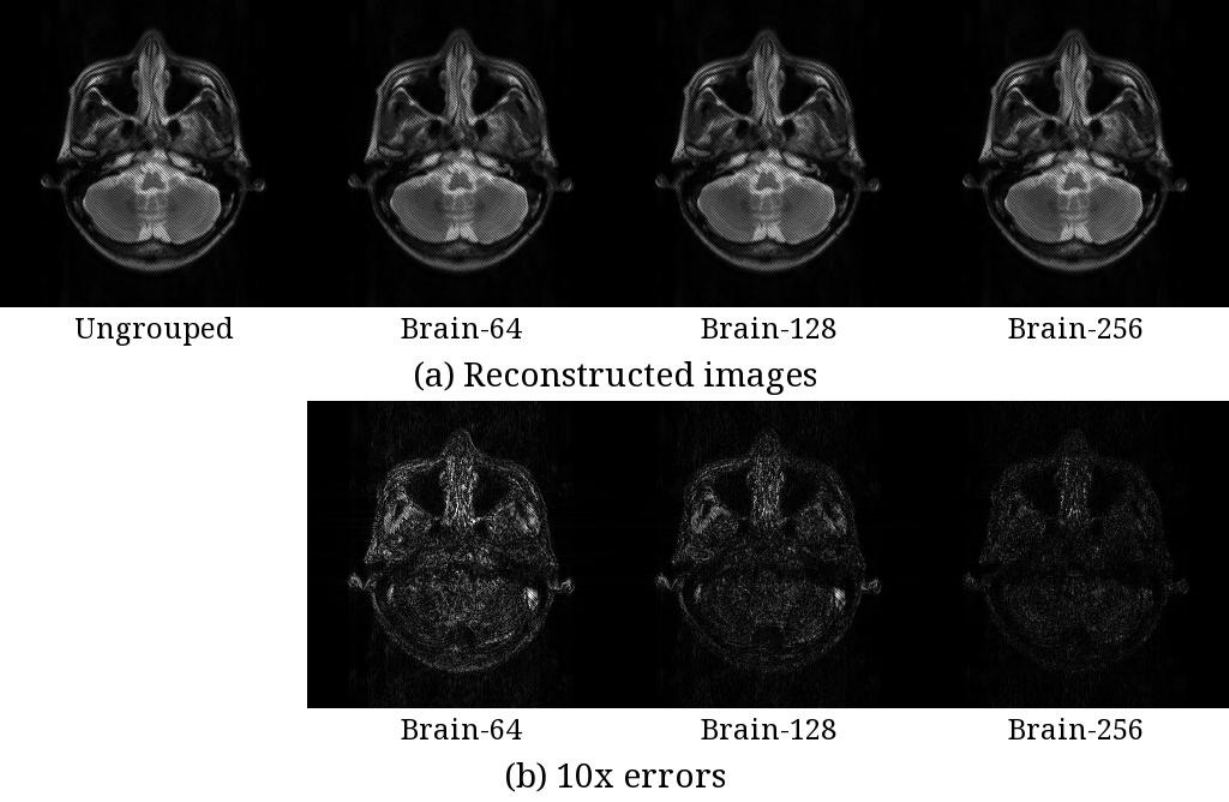}
}%
Figure S2.
Reconstructed images and their error images in the case of a T2W sequence.
(a) Reconstructed images using the brain phantom with reference (ungrouped),
64 clusters, 128 clusters and 256 clusters.
(b) Errors between the reference and grouped images.
The errors were multiplied by 10.
As shown in the error images, the image with 256 clusters
were visually close to the reference image. Abbreviations: T2W, T2-weighted.
\end{figure}

\begin{figure}[ht]%
{\centering%
\includegraphics[width=\linewidth]{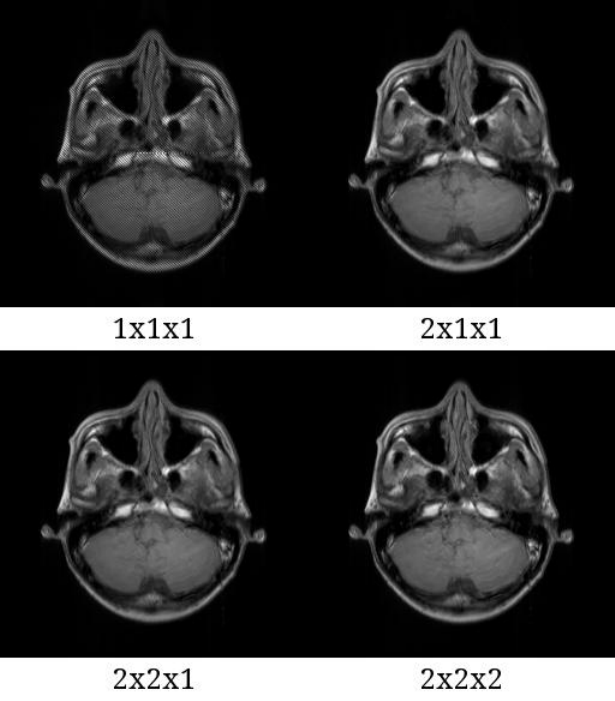}
}%
Figure S3.
Reconstructed images with 4 subvoxel types in the case of a T1W sequence.
The three numbers of the types represent the numbers of divided subvoxels
in x, y, and z directions.
Noticeable artifacts were generated at the center of the 1x1x1 image.
Abbreviations: T1W, T1-weighted.
\end{figure}

\begin{figure}[ht]%
{\centering%
\includegraphics[width=\linewidth]{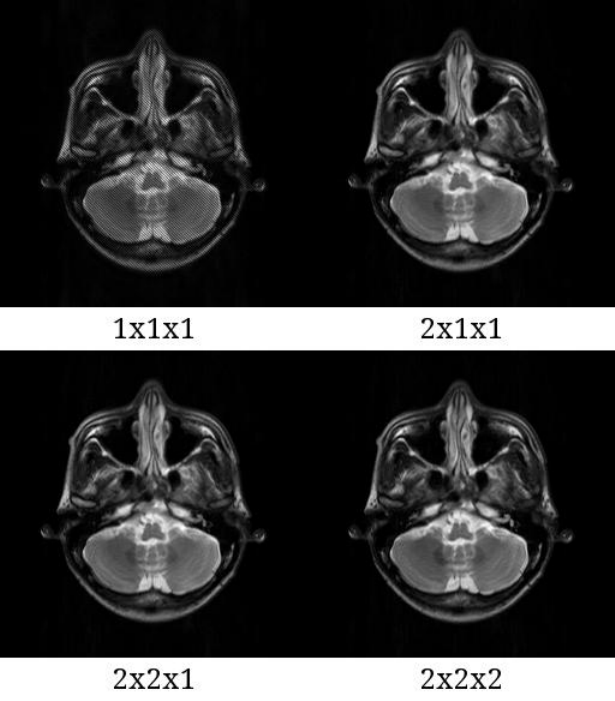}
}%
Figure S4.
Reconstructed images with 4 subvoxel types in the case of a T2W sequence.
The three numbers of the types represent the numbers of divided subvoxels
in x, y, and z directions.
Noticeable artifacts were generated in the white regions of the 1x1x1 image.
Abbreviations: T2W, T2-weighted.
\end{figure}

\begin{figure}[ht]%
{\centering%
\includegraphics[width=\linewidth]{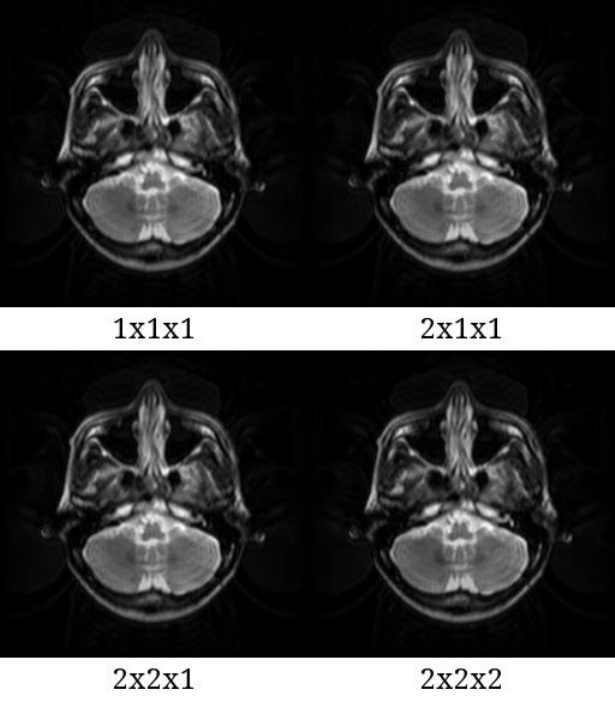}
}%
Figure S5.
Reconstructed images with 4 subvoxel types in the case of an EPI sequence.
The three numbers of the types represent the numbers of divided subvoxels
in x, y, and z directions.
The images were reconstructed with a gridding algorithm.
Abbreviations: EPI, echo-planar imaging.
\end{figure}

\begin{figure}[ht]%
{\centering%
\includegraphics[width=\linewidth]{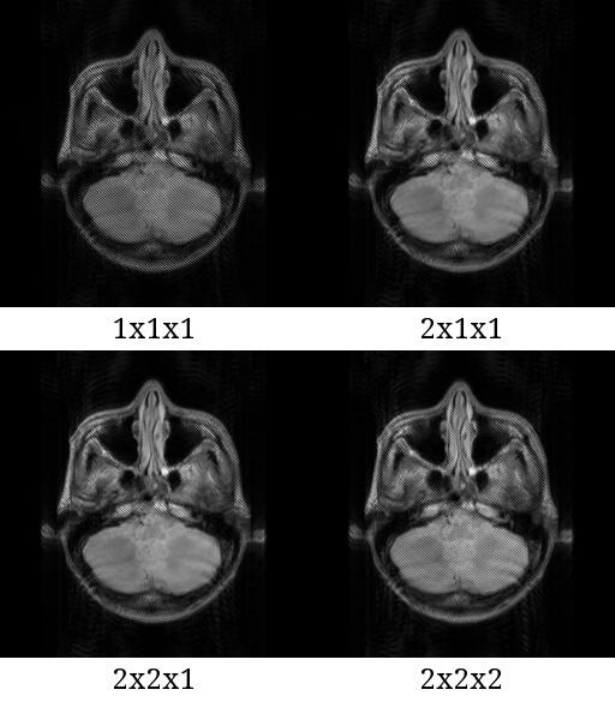}
}%
Figure S6.
Reconstructed images with 4 subvoxel types in the case of an FSE sequence.
The three numbers of the types represent the numbers of divided subvoxels
in x, y, and z directions.
There were strong artifacts in the 1x1x1 image.
The artifacts were reduced when the number of divided subvoxels was increased.
Abbreviations: FSE, fast spin echo.
\end{figure}

\end{document}